\documentclass[12pt]{iopart}
\usepackage{amssymb,hyperref}

\usepackage{graphicx}
\usepackage{subfigure}

\begin{document}

\title[Feedback and Fluctuations in a TASEP...]%
{Feedback and Fluctuations in a Totally Asymmetric Simple Exclusion Process
with Finite Resources}
\author{L. Jonathan Cook and R. K. P. Zia}
\address{Department of Physics, Virginia Tech, Blacksburg, VA 24061, USA}
\date{\today }

\begin{abstract}
\noindent We revisit a totally asymmetric simple exclusion process (TASEP) with open boundaries and a global constraint on the total number of particles [Adams, et. al. 2008 \textit{J. Stat. Mech. } P06009]. In this model, the entry rate of particles into the lattice depends on the number available in the reservoir. Thus, the total occupation on the lattice feeds back into its filling process. Although a simple domain wall theory provided reasonably good predictions for Monte Carlo simulation results for certain quantities, it did not account for the fluctuations of this feedback. We generalize the previous study and find dramatically improved predictions for, e.g., the density profile on the lattice and provide a better understanding of the phenomenon of "shock localization."
\end{abstract}

\ead{\mailto{lacook1@vt.edu},\mailto{rkpzia@vt.edu}}




\section{Introduction}

The Totally Asymmetric Simple Exclusion Process (TASEP) is a simple, yet
rich model in the poorly understood and vast realm of non-equilibrium
statistical mechanics. Particles are placed in a hypercubic lattice (for
example) and hop randomly to a nearest neighbor empty site, \emph{except }
along
one of the axes where the hopping is \emph{uni-directional}. Since its
dynamics violate detailed balance, its stationary states are non-trivial,
with typically no equivalence to any states in equilibrium statistical
mechanics. For a system with periodic boundary conditions, the stationary
state is trivial, with all configurations having equal probability (i.e.,
a flat distribution)\cite{Spitzer70}. However, the dynamic properties are
far from trivial, being quite distinct from those for a completely symmetric
exclusion processes (i.e., simple diffusion)\cite{RingDyn}. For systems with
open boundaries, in which particles hop in and out of the system with
various rates, even the stationary states are quite complex. Indeed, an open
TASEP in just one dimension settles into three different phases, depending
on the entry and exit rates\cite{Krug91}. Despite the simplicity of its
microscopic rules of evolution, the solution to this process was not known
analytically until recently\cite{Derrida92,Derrida93,Schutz93}. Needless to
say, its dynamic properties are even more complex\cite
{OpenDyn,DW,Santen02,ASZ07}. Examples of comprehensive reviews on this
``simple'' process include \cite{Derrida98,Schutz01}. Meanwhile, extensions
of such a one-dimensional TASEP are of great interest, since they may be
applied to real systems such as protein synthesis \cite{Protein}, bio-molecular motors \cite{ProMot,DW1}, 
traffic flow \cite{Traffic}, and surface growth \cite{KPZ}.

More recently, there are two studies involving a special generalization of
the open TASEP. Here, the reservoir, or ``pool,'' from which particles are
injected into the lattice is \emph{finite}. In particular, the total number
of particles on the lattice, $N$, plus the number in the pool, $N_p$, is a
fixed constant: $N_{tot}$. Such a constraint can be understood in the
context of protein synthesis as having a finite number of ribosomes (which
model particles in TASEP) in a cell\cite{ASZ08}, or in the context of
traffic (with cars as particles) as the ``parking garage problem'' \cite
{HaDenN02}. The two studies differ in how particles are moved from the
reservoir into the lattice, resulting in quite different behavior. In
particular, the former investigated the properties of $N$ and the average
particle current, $J$, when $N_{tot}$ is varied. While $J\left(
N_{tot}\right) $ displays no surprises, $N\left( N_{tot}\right) $ shows some
remarkable features (when the entry/exit rates are chosen so that, in the $%
N_{tot}\rightarrow \infty $ limit, the system is in a high-density phase or
the ``shock'' phase). Using simple arguments of self-consistency, as well as
more sophisticated domain wall (DW) theory\cite{DW,Santen02}, many of these
phenomena can be reproduced \cite{ASZ08}. In this paper, we extend this
investigation in significant ways and provide better insight into the
effects of imposing a fixed $N_{tot}$ on TASEP. In particular, we show that,
although the previous theory for $N\left( N_{tot}\right) $ appears quite
adequate everywhere, the agreement is deceptive in certain cases. Our
improvements are not merely incremental; they provide a full understanding
of the phenomenon of ``shock localization.'' As a consequence, our prediction of the average density profile, which is entirely different from that of the simple DW theory, is in excellent agreement with simulation results.

We should note that shock localization has been observed previously \cite{ProMot}.
 However, the underlying mechanisms are distinct. In particular,
the earlier studies focus on TASEP's with Langmuir kinetics, i.e., one with
no particle conservation. To model bio-moluecular motors, which can attach
and detach from a microtubule, it is natural to let particles appear and
disappear with various rates along the entire lattice. As a result, in
approaches that use DW theory and shock dynamics \cite{DW1}, the shock is
understood to be localized by a subtle interplay between adsorption and
desorption of the particles. By contrast, our study here involves a TASEP
with particle conservation, not only within the lattice, but also including
the reservoir. The shock is localized through a much simpler mechanism: the
interplay between the lattice and the reservoir. As will be shown below, the
mathematical methods used are quite similiar: site dependent hopping rates
for the domain wall.

This paper is organized as follows. In the next section, we provide the
details of the model and summarize previous results. Section \ref{DW-CT} is
devoted to both new Monte Carlo data and the improved DW theory. A summary
and outlook for future studies form the concluding section.

\section{Model definition and previous findings}

The standard open TASEP consists of a one-dimensional lattice of sites labeled
by $i=1,...,L$., each of which can be vacant or occupied by a single
particle. Thus, the configurations can be specified by the set of occupation
numbers $\left\{ n_i\right\} $, with $n_i=0,1$ being the particle content of
site $i$. Thus, $N=\Sigma _in_i$. The (random sequential) dynamics is
implemented by choosing a particle at random and moving it to the next site
with unit rate, provided the target site is not occupied. For a particle at
site $L$ (the right boundary), it leaves the system with rate $\beta $. At
the left boundary, a particle can enter the system at site $1$ with rate $%
\alpha $. When the system reaches a stationary state, the average overall
density 
\[
\rho \equiv \left\langle N\right\rangle /L
\]
will be constant. As we vary $\alpha $ and $\beta $, the system can be found
in three different phases: high density (HD), low density (LD), and maximal
current (MC). The MC phase prevails if both $\alpha $ and $\beta $ are $\geq
1/2$. In the thermodynamic limit, $\rho $ is $1/2$ and $J$, the average
current, is $1/4$, regardless of ($\alpha $,$\beta $). If $\beta <1/2$ and $%
\alpha >$ $\beta $, the system settles into a HD phase, with $\rho =1-\beta $
and $J=\beta (1-\beta )$. Due to particle-hole symmetry, the LD phase is
similar, with $\rho =\alpha $ and $J=\alpha (1-\alpha )$. for $\alpha <1/2$
and $\beta >\alpha $. The transitions across the phase boundaries HD-MC and
LD-MC are continuous. On the line $\alpha =\beta $, a HD region coexists
with a LD one, separated by a microscopic interface, known as the shock (a
term aptly describing a car driving from a high-speed, low-density region
into a traffic jam). The position of the shock wanders due to fluctuations,
so that the long-time \emph{average} of $\rho $ is again 1/2. Systems
displaying such co-existence are often referred to as being in the ``shock
phase'' (SP). A concise summary for standard TASEP is 
\begin{equation}
\rho _\infty \left( \alpha ,\beta \right) =\left\{ 
\begin{array}{ccc}
1/2 & \rm{MC; SP} & \alpha ,\beta \geq 1/2;\,\alpha =\beta \leq 1/2 \\ 
\alpha  & \rm{LD} & \alpha \leq \min \left( \beta ,1/2\right)  \\ 
1-\beta  & \rm{HD} & \beta \leq \min \left( \alpha ,1/2\right) 
\end{array}
\right.   \label{rho-STASEP}
\end{equation}
where we have used the subscript $\infty $ to remind the reader of its
relationship to our constrained system ($N_{tot}\rightarrow \infty $). The
expression for the average current is, in all cases, $\rho _\infty \left(
1-\rho _\infty \right) $.

In the constrained TASEP, the entry-rate is chosen to depend on $N_p$, the
number of particles in the reservoir. Denoting this rate by $\alpha _{eff}$,
we characterize such a dependence through a function $f$: 
\begin{equation}
\alpha _{eff}=\alpha f\left( N_p\right) .  \label{alpha-eff}
\end{equation}
The advantage of this form is that, by choosing $f\rightarrow 1$ in the
limit of large argument, we will recover the standard TASEP with the
parameters ($\alpha $,$\beta $) when the total number of particles in the
system 
\begin{equation}
N_{tot}=N_p+N  \label{N-tot}
\end{equation}
is unlimited. In \cite{HaDenN02}, the reservoir models a ``parking garage''
and the lattice models a road, so that $f$ is chosen to be the simplest
function with the desired asymptotic property: $f\left( N_p>0\right) =1$. Of
course, $f\left( 0\right) =0$, since no car can emerge from an empty garage.
By contrast, the motivation in \cite{ASZ08} is the modeling of initiation (a
ribosome attaching onto an mRNA, to begin the process of protein synthesis)
that is limited by the ribosome concentration in the cell. Thus, they
chose $f\left( x\right) \propto x$, for small $x$. Much of that study was
based on the specific function 
\begin{equation}
f\left( N_p\right) =\tanh \left( N_p/N^{*}\right)   \label{f}
\end{equation}
where $N^{*}$ models some cross-over level. As a simple starting point, it
was chosen to be the average number of particles in the standard TASEP,
i.e., $N^{*}=\rho _\infty \left( \alpha ,\beta \right) L$. Thus, 
\begin{equation}
N^{*}=\left\{ 
\begin{array}{cc}
L/2 & \alpha ,\beta \geq 1/2;\,\alpha =\beta \leq 1/2 \\ 
\alpha L & \alpha \leq \min \left( \beta ,1/2\right)  \\ 
\left( 1-\beta \right) L & \beta \leq \min \left( \alpha ,1/2\right) 
\end{array}
\right.   \label{N*}
\end{equation}
associated with the MC, LD, and HD states, respectively. As a result, it is
more convenient to express this function in terms of an intensive control
parameter 
\begin{equation}
\rho _{tot}\equiv N_{tot}/L  \label{rho_tot}
\end{equation}
and regard $f$ as a function of $\rho $. For example, for the LD case, we
have $\alpha _{eff}=\alpha \tanh \left[ \left( \rho _{tot}-\rho \right)
/\alpha \right] $. The main focus in \cite{ASZ08} are $\rho \left( \rho
_{tot};\alpha ,\beta \right) $ and $J\left( \rho _{tot};\alpha ,\beta
\right) $, with the knowledge that $\rho \left( \infty ;\alpha ,\beta
\right) =\rho _\infty \left( \alpha ,\beta \right) $ of Eqn. (\ref
{rho-STASEP}). Since predictions for $J$ follow those for $\rho $, it is
sufficient to focus only on the latter. In any case, the variations in $J$
are relatively minimal and not as spectacular as those in $\rho $.

\begin{figure}[tbh]
\par
\begin{center}
\includegraphics[height=0.45\textwidth]{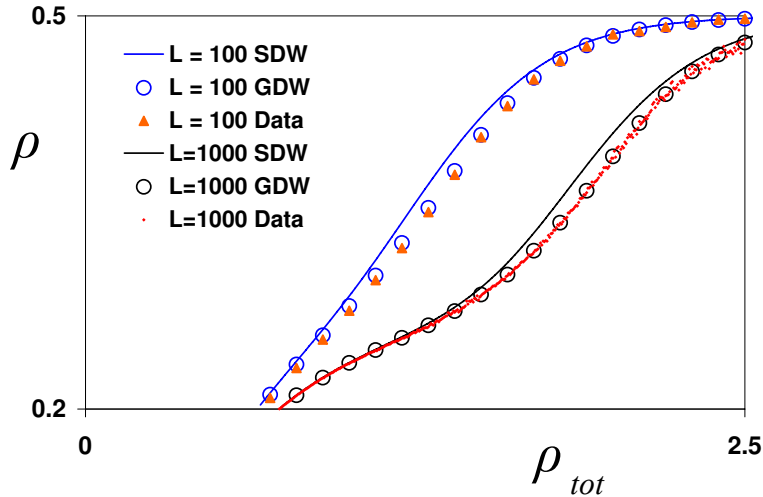}
\end{center}
\par
\vspace{-7mm}
\caption{Average overall density as a function of $N_{tot}$ for L=100 and
L=1000 with $\alpha =\beta =0.25$.}
\label{SPfig}
\end{figure}

To appreciate the different phenomena displayed by the constrained TASEP, it
is important to keep in mind that the effective entry-rate varies from $0$
to $\alpha $, as $\rho _{tot}$ is increased from $0$ to $\infty $. Thus, if $%
\alpha \ll \min \left( \beta ,1/2\right) $ (i.e., the ``LD case''), then
phase boundaries are neither traversed nor approached asymptotically. On the
other hand, if the ($\alpha $,$\beta $) are chosen so that the unconstrained
TASEP is deep in the HD or MC phase, we will cross a phase boundary, so that 
$\rho \left( \rho _{tot}\right) $ is expected to display two branches. In
the ``HD case'', there is a third branch, which reflects the presence of
coexistence, similar to the dependence of pressure on specific volume, for a
liquid-vapour system below the critical point. In this branch, $\alpha
_{eff}\cong \beta $ while $\rho $ is essentially linear in $\rho _{tot}$: $%
\rho \cong \alpha \rho _{tot}/\left( 1-\beta +\alpha \right) $. In general,
the finite size effects, by the time $L$ reaches $1000$, are hardly
noticeable. Using simple self-consistency arguments, the predicted $\rho
\left( \rho _{tot}\right) $ is in reasonably good agreement with simulation
data \cite{ASZ08}. For example, setting $\rho $ to $\alpha _{eff}$ for the
``LD case'' (in accordance with Eqn. (\ref{rho-STASEP})), we have $\rho
=\alpha \tanh \left[ \left( \rho _{tot}-\rho \right) /\alpha \right] $. This
is a transcendental equation for $\rho $, much like the one for
magnetisation as a function of the external field in the ``zeroth''
approximation of the Ising model, and leads to $\rho \left( \rho
_{tot};\alpha \right) $. The most challenging case is ``SP,'' both for
simulations and analytic understanding. 
First, the system takes a long time to settle into stationary states. 
Second, $\rho \left( \rho _{tot}\right) $
shows \emph{two} cross-overs before saturating at $1/2$. Furthermore, the
location of the second cross-over depends strongly on $L$. Meanwhile, DW
theory for the standard TASEP \cite{Santen02} accounts for some effects of
finite $L$ and provides a more detailed prediction than Eqn. (\ref
{rho-STASEP}). We denote this result of DW theory by $\rho _{DW}\left(
\alpha ,\beta ,L\right) $. Replacing $\alpha $ in this formula by $\alpha
_{eff}=\alpha \tanh \left[ 2\left( \rho _{tot}-\rho \right) \right] $,
another implicit equation for $\rho $ is established \cite{ASZ08}. Solving
such an equation leads to a $\rho \left( \rho _{tot}\right) $ that contains
all the surprising features observed. Indeed, the predicted $\rho \left(
\rho _{tot}\right) $ agrees very well with the data ($\alpha =\beta =0.25$
and $L=1000$) everywhere, except in the middle of the second cross-over,
where the worst disagreement is about 6\% \cite{ASZ08}. Unpublished data for 
$\alpha =\beta =0.25$ and $L=100$ and analytic results of $\rho \left( \rho
_{tot}\right) $ showed \cite{AZ08} that this second cross-over essentially
vanished, while the maximum disagreement there between theory and
simulations is about 7\%. In Fig. \ref{SPfig}, the solid orange symbols
represent these two sets of data, while the two lines are predictions --
with no fitting parameters -- from this ``simple'' domain wall (SDW) theory.
Given such remarkably good agreements, it may seem an overkill to pursue
this problem further. However, as we will see in the next section, the
agreement in \cite{ASZ08} is \emph{deceptively} good. Had a more sensitive
quantity like the average density \emph{profile} ($\rho _i\equiv
\left\langle n_i\right\rangle $) been considered, a glaring discrepancy
would have been exposed. 

\section{Domain wall theory for constrained TASEP}

\label{DW-CT}

The domain wall (DW) theory used in \cite{ASZ08} is formulated for a \emph{%
constant} entry (and exit) rate \cite{Santen02}, appropriate for the
standard, unconstrained TASEP. Referring the reader to the details in \cite
{Santen02}, let us summarize the key points here. The central premise of
this theory is to let the configurations of the system be approximated by
two regions of low/high densities, separated by an interface (i.e., DW) of
zero ``intrinsic width.'' To be specific, a region of low (local) density, $%
\rho _{-}$ ($=\alpha $), on sites up to $k$, is connected to a region of
high density, $\rho _{+}$ ($=1-\beta $), on sites $\in \left[ k+1,L\right] $%
, so that only a single integer ($k$) is used to label each configuration.
The dynamics of the DW is implemented through its drifting rates to the
right/left, $D_{\pm }$, dictated by both the densities on either side of the
shock and the particle fluxes from the two open ends, $j_{\pm }$. The final
result is a master equation for the probability to find the DW at site $k$
and time $t$, $P\left( k,t\right) :$%
\begin{equation}
\partial _tP\left( k,t\right) =D_{+}P\left( k-1,t\right) +D_{-}P\left(
k+1,t\right) -\left( D_{+}+D_{-}\right) P\left( k,t\right)   \label{ME01}
\end{equation}
in the bulk ($k\in \left[ 1,L-1\right] $), with 
\begin{eqnarray}
D_{+} &=&\frac{j_{+}}{\rho _{+}-\rho _{-}}=\frac{\beta \left( 1-\beta
\right) }{1-\beta -\alpha } \\
D_{-} &=&\frac{j_{-}}{\rho _{+}-\rho _{-}}=\frac{\alpha \left( 1-\alpha
\right) }{1-\beta -\alpha }
\end{eqnarray}
When the DW gets to the boundaries, it reflects back into the system, so
that 
\begin{eqnarray}
\partial _tP\left( 0,t\right)  &=&D_{-}P\left( 1,t\right) -D_{+}P\left(
0,t\right)   \label{ME02} \\
\partial _tP\left( L,t\right)  &=&D_{+}P\left( L-1,t\right) -D_{-}P\left(
L,t\right)   \label{ME03}
\end{eqnarray}
and $P\left( k<0,t\right) =P\left( k>L,t\right) \equiv 0.$ The stationary
solution is a simple exponential: 
\begin{equation}
P^{*}\left( k\right) =r^{-k}/\mathcal{N}  \label{P*0}
\end{equation}
where 
\begin{equation}
r\equiv \frac{D_{-}}{D_{+}}=\frac{\alpha \left( 1-\alpha \right) }{\beta
\left( 1-\beta \right) }  \label{r}
\end{equation}
and $\mathcal{N}$ is a normalization factor. It is clear that, for the HD/LD
phases, $r$ is greater/lesser than unity and the DW is localized at the
left/right end of the lattice. On the other hand, $r$ is unity for SP, so
that the DW can be found anywhere on the lattice.

Turning to our problem of a constrained TASEP, we recognize that, especially
if $\rho _{tot}$ is not large, the feedback from the lattice occupation
cannot be neglected. As indicated above, this is incorporated in the
previous study \cite{ASZ08} by introducing an $\alpha _{eff}\left( \rho
\right) $ and solving a self consistent equation for $\rho $. But this $\rho 
$ is just the average overall density, so that the changes in $\alpha _{eff}$
during a simulation run are completely unaccounted for. Here, we wish to
investigate how important are the density fluctuations, given that they feed
back into $\alpha _{eff}$. To distinguish the fluctuating entry rate
from the \emph{constant} in the SDW theory, let us denote the latter by $%
\alpha _{eff}^{SDW}$, so that we have explicitly, 
\begin{equation}
P_{SDW}^{*}\left( k\right) \propto r_{eff}^{-k}\,;\quad \quad
r_{eff}=\left. \alpha _{eff}^{SDW}\left( 1-\alpha _{eff}^{SDW}\right)
\right/ \beta \left( 1-\beta \right) \,\,.  \label{P*SDW}
\end{equation}

Now, to account for how this feedback affects the diffusion of the shock in
full is non-trivial. However, the essentials of the physics are clear: The
feedback stabilizes the DW, even in the vicinity of SP: If the DW wanders
too far to the right (larger $k$), there will be more particles in the
pool, resulting in a higher entry-rate. This in turn enhances the flux $j_{-}
$ and drives the DW to the left (smaller $k$). A similar stabilizing action
occurs if the DW wanders too far to the left. To implement this idea and
formulate a ``generalized'' domain wall (GDW) theory, we will make a drastic
approximation (neglecting time delays, etc.), consisting of three
ingredients:

\begin{enumerate}
\item  replacing $\alpha $ by $\alpha _{eff}$ everywhere

\item  substituting $\rho =\rho _{-}\left( k/L\right) +\rho _{+}\left(
1-k/L\right) $ into $\alpha _{eff}\left( \rho \right) $

\item  letting $\rho _{-}$ be $\alpha _{eff}$ and obtaining a $k$-dependent 
$\alpha _{eff,k}$ through a self-consistent equation.
\end{enumerate}

\noindent   
The last point is subtle and deserves clarification. As $\alpha
_{eff}=\alpha f\left( N_p\right) $, i.e., $\alpha \tanh \left[ \left( \rho
_{tot}-\rho \right) /\alpha \right] $ here, we have an implicit equation for
determining $\alpha _{eff,k}:$ 
\begin{eqnarray}
\alpha _{eff} &=&\alpha \tanh \left[ \left( \rho _{tot}-\rho _{-}\left(
k/L\right) -\rho _{+}\left( 1-k/L\right) \right) /\alpha \right] \qquad
\Rightarrow \\
\alpha _{eff,k} &=&\alpha \tanh \left[ \left( \rho _{tot}-\alpha
_{eff,k}\left( k/L\right) -\left( 1-\beta \right) \left( 1-k/L\right)
\right) /\alpha \right]
\end{eqnarray}
Once $\alpha _{eff,k}$ is found, it will enter in defining the $k$-dependent
drift rates: 
\begin{eqnarray}
D_{+,k}& =\frac{\beta (1-\beta )}{1-\beta -\alpha _{eff,k}} \\
D_{-,k}& =\frac{\alpha _{eff,k}(1-\alpha _{eff,k})}{1-\beta -\alpha _{eff,k}}%
.
\end{eqnarray}

The result of these considerations is a generalized master equation for $%
P\left( k,t\right) $ which, in the ``bulk,'' reads 
\begin{equation}
\partial _tP(k)=D_{+,k-1}P(k-1)+D_{-,k+1}P(k+1)-(D_{+,k}+D_{-,k})P(k)
\label{ME1}
\end{equation}
(with the variable $t$ suppressed). For the boundary equations, there is a
further complication which we must account for. When the resources are so
limited that the lattice cannot be fully filled at density $\rho _{+}$,
i.e., when $N_{tot}<\rho _{+}L$, the DW cannot be located arbitrarily far to
the left. Thus, the interval available to $k$ is $\left[ k_{\min },L\right] $%
, where 
\begin{equation}
k_{\min }=L-\frac{N_{tot}}{1-\beta }\rm{.}  \label{kmin}
\end{equation}
Note that $k=k_{\min }$ corresponds to the configuration with $N_p=0$, and
so, $\rho _{-}=\alpha _{eff}\left( 0\right) =0$. With this limitation in
mind, we obtain the boundary terms for the master equation: 
\begin{eqnarray}
\partial _tP(k_{\min })& =D_{-,k_{\min }+1}P(k_{\min }+1)-D_{+,k_{\min
}}P(k_{\min })  \label{ME2} \\
\partial _tP(L)& =D_{+,L-1}P(L-1)-D_{-,L}P(L)  \label{ME3}
\end{eqnarray}

Despite the extra complications, the stationary distribution can be found
analytically\cite{DW1}. Since the configuration space is one-dimensional, we have, in
general, a recursion relation $P^{*}(k)=\left[ D_{-,k+1}/D_{+,k}\right]
P^{*}(k+1)$. Thus, we arrive at the solution for our constrained TASEP: 
\begin{equation}
P^{*}(k)\propto \prod_{l=k}^{L-1}\frac{D_{-,\ell +1}}{D_{+,\ell }}
\end{equation}
and, with proper normalization, 
\begin{equation}
P^{*}(k)=\left( \sum_{m=k_{\min }}^{k-1}\prod_{\ell =m+1}^k\frac{D_{-,\ell }%
}{D_{+,\ell -1}}+1+\sum_{m=k+1}^L\prod_{\ell =k+1}^m\frac{D_{+,\ell -1}}{%
D_{-,\ell }}\right) ^{-1}  \label{P*}
\end{equation}
Since this $P^{*}(k)$ carries information on whether the DW is located
before or after a given site $i$, we can compute the average density
profile: 
\begin{equation}
\rho _i=\sum_{k=k_{\min }}^i(1-\beta )P^{*}(k)+\sum_{k=i+1}^L\alpha
_{eff,k}P^{*}(k)
\end{equation}
and of course, the overall density is given by $\rho =\sum \rho _i/L$.
Needless to say, these results are quite different from those in the simple
DW theory, e.g., Eqn. (\ref{P*0}).

\begin{figure}[tbh]
\par
\begin{center}
\includegraphics[height=0.45\textwidth]{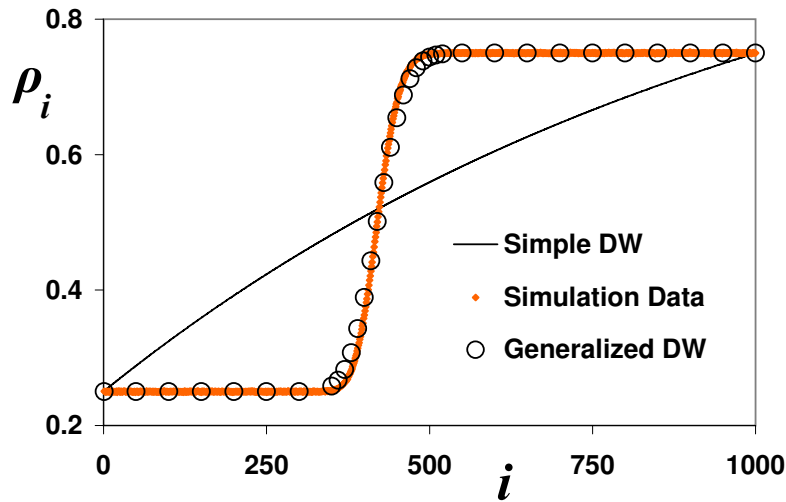}
\end{center}
\par
\vspace{-4mm}
\caption{Average density profile for $L=1000$ with $N_{tot}=800$, $\alpha
=0.75$, and $\beta =0.25$.}
\label{HDProf}
\end{figure}

To highlight how differently the two predictions compare with simulation 
data, we choose just one point in parameter space --  a point in the
``third branch'' of $\rho \left( \rho _{tot}\right) $ in the HD case. Here,
the essential physics revealed by simulations is that the number of
particles in the pool remains more or less constant as $N_{tot}$ is
increased, with the extra particles being absorbed by the lattice to form a
high density region near the exit. Furthermore, the shock on the lattice is
seen to be quite localized, as revealed by the average density profile. Fig. 
\ref{HDProf} shows this behavior for the case of $L=1000$, $N_{tot}=800$, $%
\alpha =0.75$, and $\beta =0.25$. In this figure, we also plot the
predictions from the two theories, Eqns. (\ref{P*0},\ref{P*}). Though both
provide quite good agreement with the overall density (area under the
profile), it is clear that shock localization cannot be achieved by using an
exponential $P^{*}(k)$. By contrast, accounting for the feedback through
site-dependence drift rates for the DW\cite{DW1}, the shock is localized. Given how crude our approximations are, it is remarkable that the agreement with data is so good. We conclude that even this simple-minded level of accounting for
feedback can capture the essence of shock localization. 

\begin{figure}[htb!]
\par
\begin{center}
\includegraphics[height=0.45\textwidth]{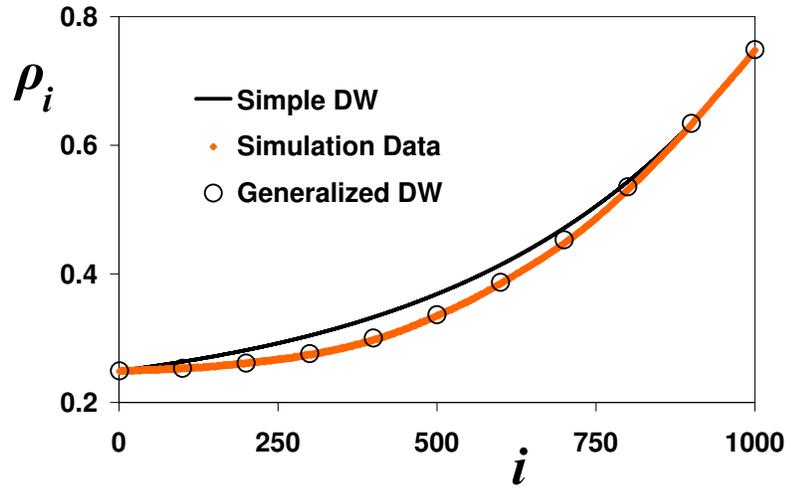}
\end{center}
\par
\vspace{-4mm}
\caption{Average density profile for $L=1000$ with $N_{tot}=2000$, $\alpha
=\beta =0.25$.}
\label{SPProf}
\end{figure}

Apart from the dramatic improvements to the profile here, our GDW theory
also provided better fits to both $\rho \left( \rho _{tot}\right) $ and the
profile in the SP case. Since our theory accounts for some feedback, we are
not surprised that it is more successful at dealing with the large
fluctuations associated with the SP case. Carrying out the computations
detailed above, we find results that are in surprisingly good agreement with
simulation data, as the open circles in Figs. \ref{SPfig} and \ref{SPProf}
show. While we show two cases in the former ($L=1000$ and $L=100$), the
latter contains only the profile for one case -- $\rho _{tot}=2$ -- in the $%
L=1000$ samples. Although the improvements over SDW here are not as
spectacular as in the HD case, they are still quite substantial.

Another useful perspective on the difference between the SDW and GDW
theories is the following. Consider the gradient of the density profile
(which is intimately related to $P^{*}(k)$, of course). For a discrete
lattice, this is 
\[
\Delta \rho _k\equiv \rho _k-\rho _{k-1}\,\,. 
\]
In SDW, it is simply $(1-\beta -\alpha _{eff}^{SDW})P_{SDW}^{*}(k)$, and so,
is a pure exponential, since $\alpha _{eff}^{SDW}$ is a constant. In other
words, $\ln \Delta \rho _k$ is linear in $k$ (with coefficient $\ln r_{eff}$%
). By contrast, in GDW, an extra $k$ dependence appears through $\alpha
_{eff,k} $ in $(1-\beta -\alpha _{eff,k})P^{*}(k)$. From Fig. \ref{HDProf},
it is clear that the non-linear terms are significant and $\Delta \rho _k$
is close to a Gaussian. On the other hand, it is also clear that Fig. \ref
{SPProf} shows that $\ln \Delta \rho _k$ is essentially linear. That GDW can
account for such sharp differences is remarkable. At a deeper level, to 
\emph{understand} how a simple $k$ dependence in $\alpha _{eff,k}$ can be so
successful will require a thorough analysis of the detailed properties of 
$f\left( N_p\right) $. Beyond the scope of this paper, this study is being
undertaken and will be reported elsewhere.

\section{Summary and outlook}

In this paper, we revisit the constrained totally asymmetric simple
exclusion process proposed in \cite{ASZ08}. Instead of particles
entering/exiting the lattice from/to an infinite pool, this system has a
fixed and finite total number of particles, $N_{tot}$ -- the sum of those on
the lattice ($N$) and in the pool ($N_p$). Further, the entry rate $\alpha
_{eff}$ is assumed to be a simple function of $N_p$, in such a way that $%
\alpha _{eff}$ $\propto N_p$ if the pool has few particles, yet saturating
at a constant $\alpha $ for $N_p\rightarrow \infty $. The main interest is
how $N$ varies as $N_{tot}$ is increased, when the exit rate is always fixed
at $\beta $. In \cite{ASZ08}, Monte Carlo simulation data showed that $%
N\left( N_{tot};\alpha ,\beta \right) $ displays a variety of properties,
depending on the parameters ($\alpha ,\beta $). A simple domain wall theory
was shown to provide good predictions (\emph{zero} adjustable parameters)
for all the data. However, there were lingering doubts for the rationale
behind such a simple theory. Here, we undertake to improve upon these
theoretical considerations.

By taking into account the fluctuating feedback into the entry rates from
the total particle occupation on the lattice, we formulated a
``generalized'' domain wall theory, in which the hopping rates for the
domain wall depend on the location of the wall (similar to the spirit,
but differing in the physics of earlier studies \cite{DW1}). Although
 this approach is still somewhat heuristic, its predictions for $N\left( N_{tot};\alpha ,\beta
\right) $ are better than those in \cite{ASZ08}. Furthermore, we find that
the agreement in \cite{ASZ08} turns out to be \emph{deceptively }good for
certain cases. In such situations, the previous predictions for a more
detailed quantity -- the density profile -- can be seriously erroneous. By
contrast, the generalized domain wall theory provides excellent results for
the profiles in all cases. In physically understandable terms, the feedback
mechanism serves to localize the domain wall. While this feedback plays a
small role in most cases, its effects are all-important in other cases, with
the most dramatic example displayed in Fig. \ref{HDProf}. Our conclusion is
that even a naive inclusion of the feedback improves significantly our
understanding of the fluctuations in a constrained TASEP. Of course, further
progress can be made, such as a more systematic approach to accounting for
all the fluctuations in this system. In particular, the quantitative 
aspects of this feedback are clearly associated with the details of $f$, specifically, its derivatives.  Beyond static properties, there are
undoubtedly many interesting dynamic phenomena yet to be discovered. For
example, we are aware of non-trivial effects\cite{A06} induced by the
constraint on the power spectrum associated with the time trace of $N$ \cite
{ASZ07} -- a phenomenon that may be understood through an appropriate 
extension of this theory.

Finally, we should remark that one of the motivations for studying TASEP
with finite resources comes from the potential applications to protein
synthesis in a cell. In a real biological system, the supply of ribosomes
(which are modeled by particles in TASEP) is finite. In case this supply is
low, there may be observable consequences for translation. For that
application, the above TASEP needs to be generalized, to include exclusion
at a distance (particles covering more than one site) and inhomogeneous
hopping rates (associated with a non-trivial sequence of codons) \cite
{Protein}. Furthermore, there are many different genes, as well as many
copies of each. Thus, we would face a system of multiple copies of TASEP's,
with different lengths and hopping rates. How these systems are affected by
finite resources will pose many interesting new challenges, especially on
the theory front. It is clear that our study here is but a very small
step towards the understanding of protein synthesis 
\emph{in vivo}. \\ 

\emph{Acknowledgements. }We thank David Adams, Jiajia Dong, and Beate
Schmittmann for unpublished results, many illuminating discussions, and a 
critical reading of the manuscript. This work
was supported in part by the National Science Foundation through DMR-0705152.
\\

\end{document}